\documentclass[twocolumn,amsmath,amssymb,10pt,superscriptaddress,a4paper,letterpaper,fleqn]{revtex4-1}
\usepackage{amssymb}
\usepackage{epsfig}
\usepackage{graphicx}
\usepackage{dcolumn}
\usepackage{array}
\usepackage{bm}
\usepackage{fancyheadings}
\usepackage{longtable}
\usepackage{multirow}
\usepackage{float}
\usepackage{afterpage}  
\usepackage{color}

\usepackage{enumitem}
\usepackage{amsmath}

\setlongtables
\usepackage[breaklinks=true,linkbordercolor={1 1 1}]{hyperref}

\begin{document}
\setcounter{page}{1}

\title{
\qquad \\ \qquad \\ \qquad \\  \qquad \\  \qquad \\ \qquad \\ 
Preequilibrium Emission of Light Fragments in Spallation Reactions}

\author{Leslie M. Kerby}
\email[Corresponding author, electronic address:\\ ]{lkerby@lanl.gov}
\affiliation{University of Idaho, Moscow, ID 83844, USA} 
\affiliation{Los Alamos National Laboratory, Los Alamos, NM 87545, USA}

\author{Stepan G. Mashnik}
\affiliation{Los Alamos National Laboratory, Los Alamos, NM 87545, USA}

\author{Arnold J. Sierk} 
\affiliation{Los Alamos National Laboratory, Los Alamos, NM 87545, USA} 

\date{\today} 

\begin{abstract}
{The ability to describe production of light fragments (LF) is 
important for many 
applications, such as cosmic-ray-induced 
single event upsets (SEUs), radiation protection, and cancer therapy 
with proton and heavy-ion beams. The cascade-exciton 
model (CEM) and the Los Alamos version of the quark-gluon string model (LAQGSM)
event generators in the LANL transport code MCNP6, describe quite well 
the spectra of fragments with sizes up to $^{4}$He across a broad range 
of target masses and incident energies (up to $\sim 5$~GeV for CEM and up 
to $\sim 1$~TeV/A for LAQGSM). However, they do not predict the high-energy 
tails of LF spectra heavier than $^4$He well. Most LF with energies above 
several tens of MeV are emitted during the precompound stage of a reaction. 
The current versions of 
our event generators do not account 
for precompound emission of LF larger than $^{4}$He. The aim of our work 
is to generalize the precompound model to include such processes, leading to 
increased predictive power of LF production. Extending the model in
this way provides preliminary results that have much better agreement 
with experimental data. 
}
\end{abstract}
\maketitle


\lhead{ND 2013 Article $\dots$}
\chead{NUCLEAR DATA SHEETS}
\rhead{L. Kerby \textit{et al.}}
\lfoot{}
\rfoot{}
\renewcommand{\footrulewidth}{0.4pt}

\section{INTRODUCTION}

Emission of light fragments (LF) from nuclear reactions is an 
interesting open question. 
Different reaction mechanisms contribute to their production; the relative 
roles of each, and how they change with incident energy, mass number of 
the target, and the type and emission energy of the fragments is not 
completely understood.

None of the available models are able to accurately predict emission of LF 
from arbitrary reactions. However, the ability to describe production of LF 
(especially at energies $\gtrsim 30$~MeV) from many reactions is important 
for different applications, such as cosmic-ray-induced single event upsets 
(SEUs), radiation protection, and cancer therapy with proton and heavy-ion 
beams. The cascade-exciton model (CEM) \cite{CEMModel, Trieste2008} 
version 03.03 and the Los Alamos version of the quark-gluon string 
model (LAQGSM) \cite{Trieste2008, LAQGSM} version 03.03 event 
generators in the Monte-Carlo n-particle transport code version 6 
(MCNP6) \cite{MCNP6}, describe quite well the spectra of fragments with 
sizes up to $^{4}$He across a broad range of target masses and incident 
energies (up to $\sim 5$~GeV for CEM and up to $\sim 1$~TeV/A for LAQGSM). 
However, they do not predict well the high-energy tails of LF spectra heavier 
than $^4$He. Most LF with energies above several tens of MeV are 
emitted during the precompound stage of a reaction. The current versions 
of the CEM and LAQGSM event generators do not account for precompound 
emission of these heavier LF. 

The aim of our work is to extend the precompound model in the codes to include 
such processes, leading to an increase of predictive power of LF-production 
in MCNP6. This entails upgrading the modified exciton model currently used 
at the preequilibrium stage in CEM and LAQGSM. It will also include expansion 
and examination of the coalescence and Fermi break-up models used in the 
precompound stages of spallation reactions within CEM and LAQGSM. Extending 
our models to include emission of fragments heavier than $^4$He at the 
precompound stage already gives preliminary results with much
better agreement with experimental data.

\section{THEORETICAL BACKGROUND}

Our models consider that
a reaction begins with the IntraNuclear Cascade, referred to as 
the INC.
The incident particle or nucleus 
(in the case of LAQGSM) enters the target nucleus and begins 
interacting with nucleons, scattering off them and also often creating 
new particles in the process. The incident particle and all newly 
created particles are followed until they either escape from the 
nucleus, are absorbed, or, for nucleons, reach a threshold energy
(roughly 10-30 MeV) 
and are then considered ``absorbed" by the nucleus. 

The preequilibrium stage uses the modified exciton model (MEM) to 
determine emission of protons, neutrons, and fragments up to $^4$He 
from the residual nucleus. We discuss the MEM in more detail below. 
This stage can have a highly excited residual nucleus undergoing 
dozens of exciton transitions and particle emissions. The 
preequilibrium stage ends when the residual nucleus is just as 
likely to have a $\Delta n = +2$ exciton transition as a 
$\Delta n = -2$ exciton transition.

In the evaporation stage, neutrons and protons in
the residual nucleus can ``evaporate," either singly or as fragments. 
The CEM evaporation stage is modeled with
a modification of the Furihata's generalized 
evaporation model code (GEM2) \cite{GEM2}, and includes light fragments 
up to $^{28}$Mg.

During and after evaporation, the code looks to see if we have an 
isotope that has $Z \geq 65$ and is fissionable. If it is, and there 
is fission, then the code also allows evaporation from the fission 
fragments.

There are two models that are not directly parts of this 
progression: coalescence and Fermi break-up. The INC 
stage only emits neutrons, protons, and pions (and other particles, 
when using LAQGSM at high energies), so the coalescence 
model ``coalesces" some of the INC neutrons and protons 
into larger fragments, by comparing their momenta. If their 
momenta are similar enough then they coalesce. The current coalescence 
model can only coalesce up to $^4$He fragments, the same as the 
preequilibrium stage. The Fermi break-up is an oversimplified 
multifragmentation model that is fast and accurate for small atomic 
numbers; in the current CEM model it is used when any residual nucleus
or fragment has a mass number less than 13.

The 
MEM used by CEM and LAQGSM \cite{CEMModel, Trieste2008, LAQGSM} 
calculates $\Gamma _j$, the emission width (or probability of 
emitting particle fragment $j$) as
\begin{equation}
\Gamma_{j}(p,h,E) = \int_{V_j^c}^{E-B_j} \lambda_c^j (p,h,E,T)dT ,
\end{equation}
where the partial transmission probabilities, $\lambda_c^j$, are equal to
\begin{eqnarray}
\lambda_c^j(p,h,E,T) = \frac{2s_j + 1}{\pi^2\hbar^3} \mu_j \Re (p,h) \times \nonumber \\
\times  \frac{\omega (p-1,h,E-B_j-T)}{\omega (p,h,E)} T \sigma_{inv} (T) .
\label{LambdaTransmission}
\end{eqnarray}
Eq.~(\ref{LambdaTransmission}) describes the emission of neutrons 
and protons. For complex particles, the nuclear level density $\omega$ 
becomes more complicated and an extra phase-space factor $\gamma_j$
must be introduced:
\begin{equation}
\gamma_j \approx p_j^3 (\frac{p_j}{A})^{p_j - 1} .
\label{GammaBeta}
\end{equation}

Eq.~(\ref{GammaBeta}) for $\gamma_j$ is a 
only a rough 
estimation that we improve by parameterizing it over a mesh of residual 
nuclei energy and mass numbers in our codes \cite{CEMUserManual}. 
As the MEM uses a Monte-Carlo technique to solve the master equations
describing
the behavior of the nucleus at the preequilibrium stage (see details in 
\cite{CEMModel}), it is very easy to extend the number of types of possible
LF that can be emitted during this stage. We generalize the MEM to consider
the possiblity of emission up to 66 types of nucleons and LF, up to $^{28}$Mg. 
As a starting point, for the inverse cross sections, Coulomb barriers, and
binding energies
of all LF we use the approximations adopted by GEM2 \cite{GEM2}.

\section{RESULTS}

Extending the Fermi break-up model to include heavier LF (up to $A = 16$) 
allows us to use it for nuclei with $A > 12$, yielding
increased accuracy for reactions with light targets. Below are 
examples of calculations by CEM (with the expanded Fermi
breap-up model) 
compared to experimental data \cite{Hagiwara,Uozumi}.

\begin{figure}[!htb]
\includegraphics[width=0.9\columnwidth]{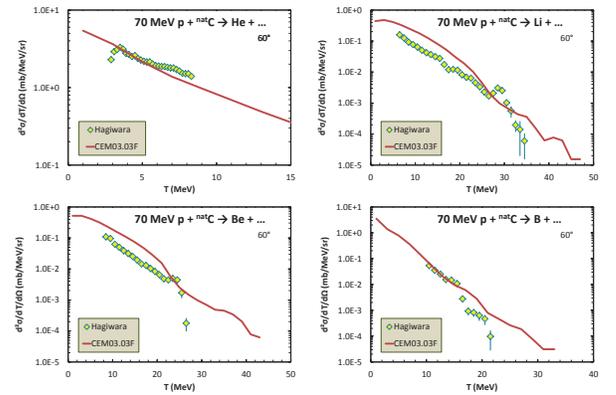}
\caption[]{Comparison of CEM03.03F results(solid red lines;
``F'' stands for expanding the range of LF in CEM) 
with experimental data 
by Hagiwara et al. \cite{Hagiwara} (open symbols) for 70 MeV protons
on a natural carbon target. 
We calculate for $^{12}$C only.}
\label{Hagiwara}
\end{figure}

\begin{figure}[!htb]
\includegraphics[width=0.9\columnwidth]{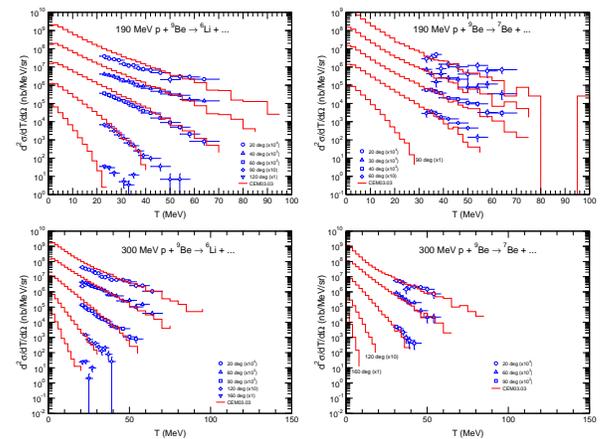}
\caption[]{Comparison of CEM03.03F results (solid red lines) with
experimental data 
by Uozumi et al. \cite{Uozumi} (open symbols) for 190 MeV protons 
on a $^9$Be target.}
\label{Uozumi}
\end{figure}

Results from the extended Fermi break-up model achieve
good agreement with experimental results for these light targets.

\begin{figure}[!h]

\vspace*{-4mm}
\centering
\includegraphics[width=0.6\columnwidth]{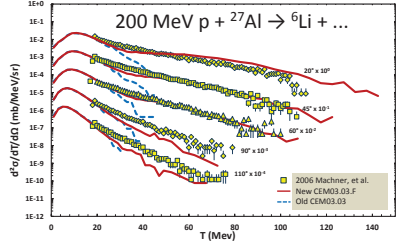}
\vspace*{-1mm}
\caption[]{Comparison of CEM03.03F results (solid red lines) with experimental 
data by Machner et al. \cite{Machner} (open symbols).}
\label{p200AlLi}
\end{figure}

\begin{figure}[!h]

\vspace*{-4mm}
\centering
\includegraphics[width=0.6\columnwidth]{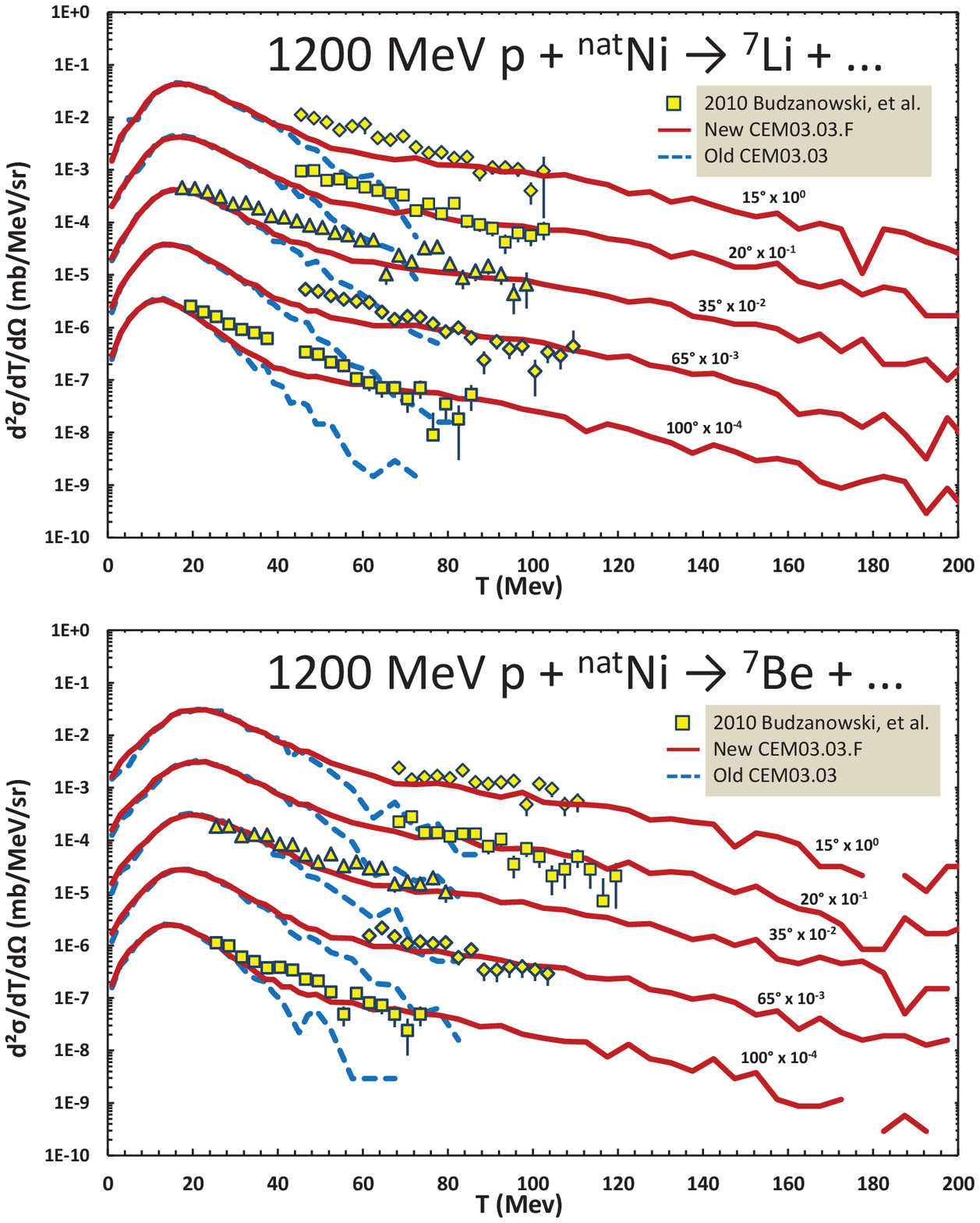}
\vspace*{-1mm}
\caption[]{Comparison of CEM03.03F results (solid red lines) withd experimental 
data by Budzanowski et al. \cite{Budzanowski} (open symbols).}
\label{p1200Ni}
\end{figure}

Expanding the MEM to include heavier LF (up to $^{28}$Mg) yields 
increased accuracy for several reactions we investigate. Figure~\ref{p200AlLi} 
compares new simulations from our expanded model with data by 
Machner et al. \cite{Machner} for 200 MeV p + $^{27}$Al.

We also find that the integral spectra for n, p, d, t, $^3$He,
and $^4$He (not shown), are not significantly impacted by this LF
emission expansion.

Figure~\ref{p1200Ni} compares our results for 1200 MeV p + $^{nat}$Ni with 
 new data by Budzanowski et al. \cite{Budzanowski}.

Similar results for different LF spectra are obtained for several
other reactions (see, e.g., \cite{Kerby2012}).

Our results indicate that expanding the MEM to include LF preequilibrium 
emission significantly increases accuracy of the high-energy spectra compared 
to experimental data.

\section{ CONCLUSIONS}

Extending the CEM model to include emission of light fragments (LF) heavier 
than $^4$He (up to $^{28}$Mg) in the preequilibrium stage results in significantly 
improved accuracy compared to experimental data for several reactions, especially in the high-energy tails of the spectra.

Future work includes finding a global parametrization for $\gamma_{\beta}$,
incorporating the expanded event generators into MCNP6, adding 
coalescence of heavier fragments, further exploring Fermi break-up,
and upgrading the evaporation model. 

This study was carried out under the auspices of the National Nuclear Security 
Administration of the U.S. Department of Energy at Los Alamos National Laboratory under Contract No. DE-AC52-06NA253996.

\end{document}